\documentclass[11pt]{proc}
\usepackage[dvips]{graphicx}

\begin{document}

\title{Analysis of Non-Gaussian Nature of Network Traffic and its
	Implication on Network Performance}

\author{Tatsuya Mori\dag,\quad\quad Ryoichi Kawahara\ddag\quad\quad
Shozo Naito\dag\\
tatsuya@nttlabs.com, \{kawahara.ryoichi, naito.shozo\}@lab.ntt.co.jp\\
\dag NTT Information Sharing Platform Laboratories, NTT corporation\\
\ddag NTT Service Integration Laboratories, NTT corporation
}

\maketitle

\begin{abstract}
	We analyzed the non-Gaussian nature of
	network traffic using some Internet traffic data.
	We found that (1) the non-Gaussian nature degrades  network
	 performance, (2) it is caused by `greedy flows' that exist with
	 non-negligible probability, and (3) a large majority of `greedy flows'
	 are TCP flows having relatively small hop counts, which correspond to
	 small round-trip times.
	 We conclude that in a network that has greedy flows with
	 non-negligible probability, a traffic controlling scheme or bandwidth
	 design that considers non-Gaussian nature is essential.

\end{abstract}

\section{Introduction}
	 Traffic characterization based on measurements is crucial for
	 establishing high-quality performance evaluation and efficient
	 network provisioning.
	 It has been widely elucidated that in today's high-speed data
	 networks, self-similarity is appropriate for traffic
	 characterization and performance evaluation since pioneering work
	 conducted by researchers from Bellcore in the early 1990s
	 \cite{willinger-94,park-00}.
	 Self-similarity in data network traffic suggests that traffic
	 variability has `long-range dependence (LRD)', while the classic
	 Poisson traffic model is based on the principle that traffic
	 variability has short-range dependence (exponential).
	 In a series of self-similarity related studies, it was found that
	 if traffic variability has a higher degree of LRD, then network
	 performance tends to be worse than when it has a lower one
	 \cite{norros-95,KGM97,park-00}.

	 Self-similarity in data networks has been studied in terms of its
	 ubiquitous presence and usefulness from various measurements and
	 statistical analyses and simulation studies. However, these are
	 not sufficient from the viewpoint of evaluating network
	 performance.
	 That is, in some cases network traffic with a higher degree of LRD
	 could show better performance than that with a lower one --- the
	 reverse of above findings.
	 This is because LRD reflects only the temporal structure of traffic
	 variability and not its spatial structure, such as marginal
	 distribution.
	 Grossglauser et al. showed, using a fluid traffic model
	\cite{gross-99}, that in addition to LRD, the difference in
	marginal distributions strongly affect network performance.
	 However, detailed information about marginal distributions of real
 network traffic and the way to describe them in their fluid traffic
 model was not given in the study.
 In general, if traffic is aggregated from a number of independent and
 identically distributed flows, the marginal distribution of its
 variability is considered to be Gaussian and this nature is
 guaranteed by the central limit theorem.
 Actually, in recent traffic models that reflect traffic LRD such as
 the well known fractional Brownian motion traffic
 model proposed by Norros \cite{norros-95} or the one proposed by
 Willinger and Taqqu \cite{willinger-97}, which is a superposition of
 a large number of independent ON/OFF sources with heavy-tailed ON
 and/or OFF periods, their marginal distribution of traffic
 variability is Gaussian.

 In this paper, we investigate marginal distributions of network
 traffic using some Internet traffic data.
 We found that marginal distributions of traffic variability are
 not always Gaussian (i.e., they are non-Gaussian) and in many
 cases, they are skewed positively.
 We also show that the non-Gaussian nature has a strong influence
 on network performance.
 This means that it is essential to consider the non-Gaussian
 nature of network traffic in order to characterize traffic
 variability.
 Thus, we focus on the mechanisms that cause non-Gaussian nature of
 traffic variability.
 To study this, we analyze the behavior of each IP flow composing
 aggregated traffic from the viewpoints of size distribution, hop
 counts, RTT, and protocol.

 This paper is organized as follows. 
 Section 2 shows examples of the non-Gaussian nature of network traffic
 and its implication on network performance.
 In section 3, to identify the mechanisms causing non-Gaussian nature, we
 define `per-time-unit flow' and analyze its statistics such as size
 distribution, hop counts, RTT, and protocol.
 Section 4 gives our conclusions.

\section{Examples of Non-Gaussian Nature and performance implication}
	This section shows examples of the non-Gaussian
	nature of network traffic using real Internet traffic traces and 
	then shows its performance implications by trace driven
	simulation.

	\subsection{Data}
	We used traces from three different sets of network traffic
	for our analyses.
	In this work, the length of all traces was set to
	300 s to avoid the effect of non-stationarity and also to get
	enough statistics. Actually, each trace had at least
	100,000 packets in this condition.
	Details about the traces are briefly summarized as follows.

		\subsubsection*{Data I: ECL external line}
		This line is the main external connection line of NTT
		R\&D center (ECL).
		It is a 12-Mbps ATM line and traces were captured
		at the segment one hop before the line.
		The measurements were made during daily busy hours on
		some weekdays in July 2001.
		In total we used 48 traces for this study.

		\subsubsection*{Data II: OCN-SINET}
		This line connects NTT's Open Computer Network (OCN)
		and the Science Information Network (SINET).
		OCN is NTT's commercial Internet backbone network and
		SINET is the largest Internet backbone network for
		scientific research institutes in Japan.
	 	The link is a 135-Mbps ATM line.
		The measurements were made during daily busy hours on
		some weekdays in January 2000.
		In total we used 34 traces for this study.
		More detailed information about this data is
		available in \cite{kawahara-01}.
  
		\subsubsection*{Data III: Bellcore}  
		The lines are several Ethernet networks at the Bellcore
		Morristown Research and Engineering facility.
		Traces are available from the Internet Traffic Archive
		\cite{ita}.
		In this study, we used the first 300 s of BC-pAug89.TL.
		Detailed information about the traces is shown in
		\cite{willinger-94}, where the self-similarity of
		Ethernet traffic was first demonstrated using data
		that included this data set.

	\subsection{Traffic variability and marginal distributions}
\begin{figure}[tbp]
	 \begin{center}
	\includegraphics[width=85mm]{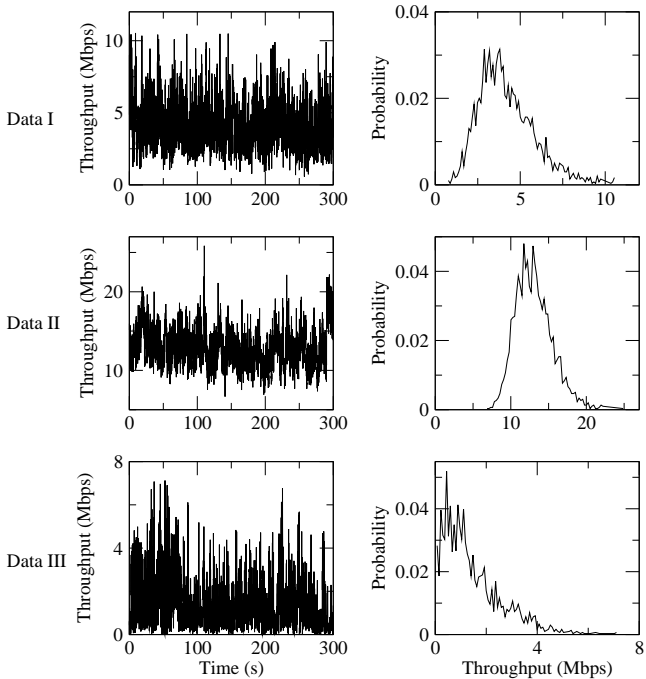}
	  \caption{Examples of non-Gaussian nature of traffic variability.}
	  \label{fig:non-Gauss-ex}
	 \end{center}
 	\end{figure}

	For all traces described in 2.1, we calculated the throughput
	variability and their marginal distributions.
	Here, we calculated throughput using a time interval of 0.1 s.
	Figure \ref{fig:non-Gauss-ex}, shows throughput variability
	(left side) and	marginal distributions (right side) for
	randomly chosen traces from the three networks described in
	2.1.
	From this figure, we can intuitively find that marginal
	distributions of all traces are asymmetric and skewed
	positively.
	To characterize the difference in marginal distributions
	quantitatively, we used skewness, which is defined as 
	\begin{equation}
	skewness =
	\frac{\left<\left(X-\left<X\right>\right)^3\right>}{\sigma^3},
	\end{equation}
	where $\left<X\right>$ is the mean of $X$ and $\sigma$ is
	the standard deviation of $X$.
	If the distribution is skewed positively (negatively),
	skewness is positive (negative). 
	If the distribution is exactly Gaussian, the shape of the
	marginal distribution is symmetric and the skewness is 0.
	Table 1 shows skewness of three example traces given in
	Fig.~\ref{fig:non-Gauss-ex}.
	All values of skewness took positive values.
	In these three example traces, we can see that traffic variability
	of Data III has the strongest non-Gaussian nature.

	\begin{table}[tbp]
	\begin{center}
	\caption[]{Skewness of three example traces.}
	\label{tab:skew-ex}
		\begin{tabular}{c|ccc}
		\hline
			   & Data I & Data II & Data III\\
		\hline
		$skewness$ & 0.812 & 0.655 & 1.320 \\
		\hline
		\end{tabular}
	\end{center}
	\end{table}

	\subsection{Non-Gaussian nature and network performance}
	For Data I and II, we performed trace-driven simulation to see the
	effect of non-Gaussian nature on network performance in conditions
	having similar bandwidth and buffer capacity.
	We used Internet-to-ECL	traffic for Data I and OCN-to-SINET
	traffic for Data II because these directions have more
	traffic than the opposite directions.
	We did not use Data III because we could not clarify the
	direction of traffic from given traces.
	
\begin{figure}[tbp]
	 \begin{center}
	\includegraphics[width=85mm]{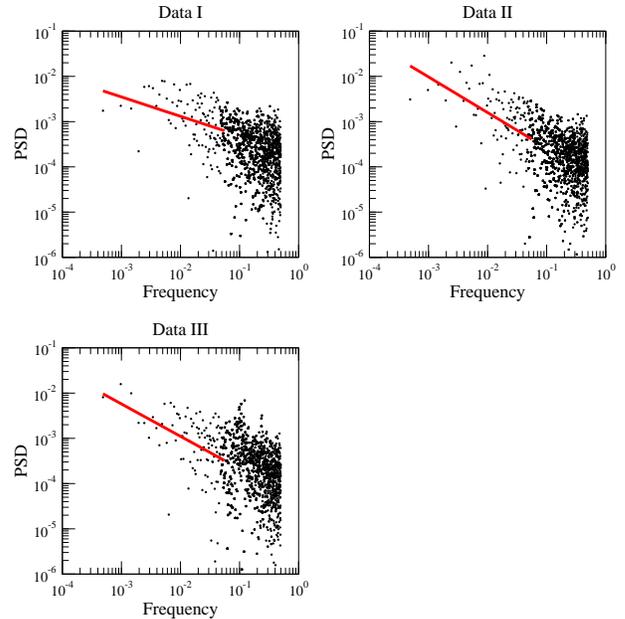}
	  \caption{Power spectrum density for three traces given in
	 Fig.~\ref{fig:non-Gauss-ex}.}
	  \label{fig:hurst-ex}
	 \end{center}
	\end{figure}

	We also show the relationship between the Hurst parameter and
	network	performance for comparison.
	The Hurst parameter, $H$, is between 0.5 and 1, where a large
	value means a high degree of LRD
	\cite{willinger-94},\cite{park-00}.
	To estimate $H$, we employed power spectrum
	density estimation, removing the linear trend from
	the throughput time series before employing a Fourier transform
	(Fig.~\ref{fig:hurst-ex}).
	From the slope $\alpha$ of the log-log regression of the power
	spectrum density versus the frequency, we get $H =
	\frac{1-\alpha}{2}$ \cite{beran},\cite{murad95}.
	Here, the lowest 10\% of frequency was used for regression.
	Table~\ref{tab:hurst-ex} shows estimated Hurst parameters of
	three example traces given in Fig.~\ref{fig:non-Gauss-ex}.
	All values are larger than 0.5, which
	indicates that all traces have LRD.

	\begin{table}[tbp]
	\begin{center}
	\caption[]{Estimated Hurst parameters of three example traces.}
	\label{tab:hurst-ex}
		\begin{tabular}{c|ccc}
		\hline
		    & Data I & Data II & Data III\\
		\hline
		$H$ &  0.714 & 0.894 & 0.858 \\
		\hline
		\end{tabular}
	\end{center}
	\end{table}

	\begin{figure}[tbp]
		\setlength{\unitlength}{0.9mm}
		\begin{center}
		\begin{picture}(90,20)
		\put(7,15){\oval(20,10)}
		\put(2,12){\shortstack{Trace\\source}}
  
		\put(30,15){\circle{10}}
		\put(30,15){\circle{11}}
		\put(26,12){\shortstack{Node\\1}}
		\put(50,15){\circle{10}}
		\put(50,15){\circle{11}}
		\put(46,12){\shortstack{Node\\2}}
		\put(73,15){\oval(20,10)}
		\put(68,12){\shortstack{Trace\\sink}}
		\put(16,15){\vector(1,0){8}}
		\put(55,15){\vector(1,0){8}}
		\put(34,15){\vector(1,0){11}}
		\put(39,8){\line(0,1){7}}
		\put(28,5){\shortstack{aggregation link}}
		\put(28,1){\shortstack{bandwidth = $Bw$}}
		\end{picture}
		\end{center}
		\caption{Simulation environment.}
		\label{fig:tracesim}
	\end{figure}
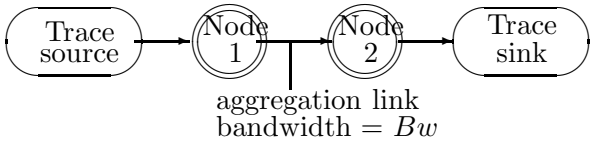  

	Figure~\ref{fig:tracesim} shows our simulation model.
	As the network simulator, we used ns-2 \cite{ns-2}.
	The one-way trace data was set to `Trace source' and
	packets were sent from it to `Trace sink' via two
	nodes following the timestamp and packet size recorded
	in the trace data.
	In this study, there was no host-side traffic control
	scheme such as TCP; all packets were treated like
	UDP packets.
	This is because we wanted to see how packets would be
	discarded given a constrained bandwidth for originally
	demanded (non-shaped) traffic.
	Here, we set the buffer size of the `aggregation link'
	to 50 packets to see the difference in performance
	clearly, and used FIFO as the packet scheduling.
	The link bandwidth was changed so that the link utilization of
	each trace was 0.6.
	For example, if the average traffic variability	(throughput) of a
	trace was 6 Mbps, we set the bandwidth to 10 Mbps.

	\begin{figure}[tbp]
	 \begin{center}
	\includegraphics[width=90mm]{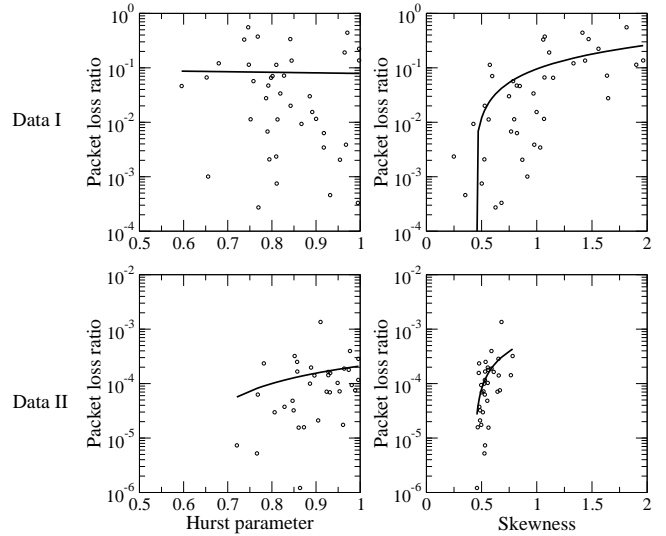}
	  \caption{Results of trace-driven simulation. Left side shows Hurst parameter vs. packet loss ratio; right side shows skewness vs. packet loss ratio. Top is the result using Data	I and bottom is Data II.}
	  \label{fig:h-s-ploss}
	 \end{center}
	\end{figure}

	\begin{table}[tbp]
	\begin{center}
	\caption[]{Correlation coefficients of Hurst parameter
	and skewness vs. packet loss ratio.}
	\label{tab:h-s-ploss}
		\begin{tabular}{c|cc}
		\hline
		 & Data I & Data II\\
		\hline
		Hurst parameter & -0.015 & 0.169 \\
		\hline
		skewness        & 0.584  & 0.435 \\
		\hline
		\end{tabular}
	\end{center}
	\end{table}

	For each trace, we calculated its skewness and Hurst parameter
	from its throughput variability, and performed the trace-driven
	simulation.
	Figure~\ref{fig:h-s-ploss} shows the results where each point
	corresponds to the result of one trace.
	From the figure, we can see that both Hurst parameter
	and skewness took a wide range of values for two networks.
	That is, most of traffic had LRD and was positively skewed
	(non-Gaussian).
	It also shows that as skewness increased, network performance
	degraded as well.
	That is, as the non-Gaussian nature became stronger, 
	the network performance became more degraded.
	To see the relationship between network performance	and the above
	two characteristics (LRD, non-Gaussian) quantitatively, we
	calculated correlation coefficients between them
	(Table~\ref{tab:h-s-ploss}).
	In Fig.~\ref{fig:h-s-ploss}, solid lines indicate
	linear regression.
	The table shows that skewness was positively correlated with
	network performance while the Hurst parameter had little
	correlation with it.
	These results imply that the non-Gaussian nature of traffic
	strongly affects the network performance.

\section{Mechanism of Non-Gaussian Nature}
	This section investigates the mechanism of the non-Gaussian nature
	of network traffic.
	For this, we first introduce `per-time-unit flow' --- an IP
	flow defined in a given time unit --- to analyze the behavior of
	IP flows composing the aggregated traffic.
	Here IP flow is a group of packets having a unique combination
	of source IP address, destination IP address, source port,
	destination port, and protocol as is defined in
	\cite{thompson-97,kawahara-01}.
	Then we show that `greedy flows' strongly affect the non-Gaussian
	nature of network traffic.
	After that, we show the nature of greedy flows from the	viewpoints
	of hop counts distributions.
	Investigating the relationship between traffic variability and hop
	count distributions will also enable us to establish efficient
	network	bandwidth design according to its topology.
	To investigate the nature of greedy flows in detail, we also
	studied	RTT and protocol distributions of per-time-unit flows.

	\subsection{Definition of per-time-unit flow and greedy flows}
	Many recent ON/OFF source traffic models (also known as
		packet train models) such as \cite{willinger-97}	and
	\cite{KGM97} assume that traffic is aggregated from a number of
	flows having a {\it uniform} rate.	
	Accordingly, each flow has a similar size on a certain time scale
	when the flow is in the ON-period.
	It should also be pointed out that in these models, aggregated
	traffic shows the {\it Gaussian} nature according to the central limit
	theorem, as mentioned in section 1.
	However, it is not clear that the above ``uniform rate assumption'' is
	appropriate for modern Internet traffic.
	So we introduce `per-time-unit flow' to investigate how traffic is
	aggregated on a certain time scale, and to see how the behavior of
	each flow contributes to the non-Gaussian nature of aggregated
	traffic.

\begin{figure}[tbp]
	 \begin{center}
	\includegraphics[width=80mm]{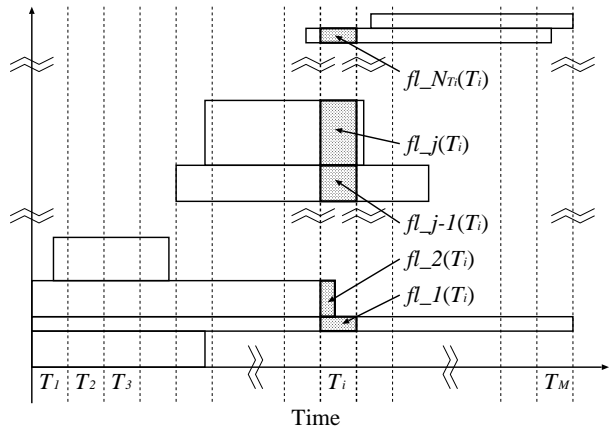}
	  \caption{Definition of per-time-unit flow.}
	  \label{fig:def-per-time-unit}
	 \end{center}
	\end{figure}

	In Fig.~\ref{fig:def-per-time-unit}, each square corresponds
	to one IP flow.
	We divided traces into time	unit $T_i$, where $1\leq i \leq M$.
	For all $i$, the length of $T_i$ was set to time interval
	$\tau$. 
	For each $T_i$, we define per-time-unit flow
	$fl\_j\left(T_i\right)$ as shown by the shaded regions in the
	figure, where  $1 \leq j \leq N_{T_i}$ and $N_{T_i}$ is the
	number of flows during $T_i$.
	The per-time-unit flow $fl\_j\left(T_i\right)$ should contain
	at least two packets during $T_i$.
	In this work the length of $\tau$ was set to 0.1 s.
        
	For each per-time-unit flow $fl\_j\left(T_i\right)$, we
	counted the number of packets
	$N_p\left(fl\_j\left(T_i\right)\right)$.
	In this study, we defined a `greedy flow' as one whose
	$N_p\left(fl\_j\left(T_i\right)\right)$ is larger than 20,
	which corresponds to throughput of about 1 Mbps assuming the
	average packet size to be 700 bytes.

	\subsection{Size distribution of per-time-unit flow}
	For Data I and II, we investigated the size distribution
	of $N_p\left(fl\_j\left(T_i\right)\right)$.
	We calculated the following complementary cumulative
	distribution for all $i,j$.

	\begin{equation}
		P\left[N_p\left(fl\_j\left(T_i\right)\right) >
		n_p\right]
	\end{equation}

\begin{figure}[tbp]
	 \begin{center}
	\includegraphics[width=85mm]{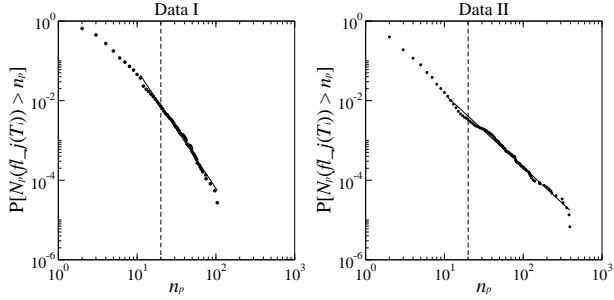}
	  \caption{LLCD plots of
	 $N_p\left(fl\_j\left(T_i\right)\right)$ for two traces given
	in Fig.~\ref{fig:non-Gauss-ex}.}
 	  \label{fig:per-time-unit-numpkt-dist}
	 \end{center}
	\end{figure}

	Figure~\ref{fig:per-time-unit-numpkt-dist} shows the log-log
	complementary cumulative distribution (LLCD) plots of
	$N_p\left(fl\_j\left(T_i\right)\right)$ for all $i,j$.
	As we can see immediately, the figure shows that distributions
	of $N_p\left(fl\_j\left(T_i\right)\right)$ are in good
	agreement with the power-law; that is,

	\begin{equation}
		P\left[N_p\left(fl\_j\left(T_i\right)\right) >
		n_p\right] \sim n_p^{-\alpha},\quad as\quad n_p \to \infty.
	\end{equation}

	Estimated power exponents $\alpha$ of equation (3) for traces
	given in Fig.~\ref{fig:non-Gauss-ex} are 2.96 for Data I and
	1.83 for Data II,
	where the regression range was set to $n_p \geq 10$.
	Correlation coefficients for regressions are -0.99 for both
	Data I and II.
	The power-law of distributions of
	$N_p\left(fl\_j\left(T_i\right)\right)$ indicates that both 
	traces described in Fig.~\ref{fig:per-time-unit-numpkt-dist}
	had greedy flows with non-negligible probability (right side of
	the dashed line in the figure.
	It should also be pointed out that as $\alpha$ approaches 2,
	the distribution of	$N_p\left(fl\_j\left(T_i\right)\right)$
	approaches a heavy-tailed
	\footnote{The distribution of $X$ is heavy-tailed
	if\\$P\left[X>x\right] \sim x^{-\alpha},\quad	as\quad x\to
	\infty,\quad 0<\alpha<2$}
	distribution, which indicates that very large values existed
	with non-negligible probability.

	\subsection{Greedy flows and non-Gaussian nature}
\begin{figure}[tbp]
	 \begin{center}
	\includegraphics[width=85mm]{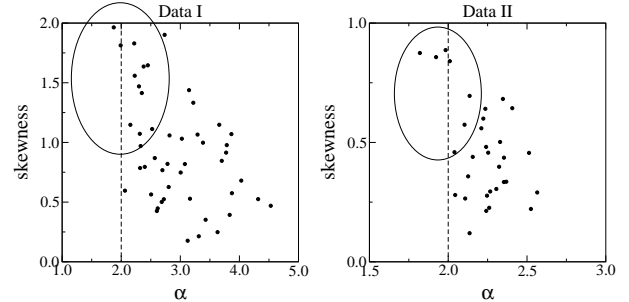}
	  \caption{Exponents $\alpha$ vs. skewness. Left side is Data
	 I and right side is Data II.}
 	  \label{fig:llcd-exp}
	 \end{center}
	\end{figure}

	In order to see the relationship between greedy flows and
	the non-Gaussian nature of network traffic, we investigated the
	relationship between estimated power exponents $\alpha$ of
	equation (3) and the skewness of throughput variability for each
	trace.
	In Fig.~\ref{fig:llcd-exp}, the estimated power
	exponents $\alpha$ are plotted against skewness for all
	traces of Data I and II.
	The figure shows that in both cases, skewness increased as
	$\alpha$ decreased, and this tendency was stronger
	when $\alpha$ was close to 2 (inside ellipses and
	dashed lines).
	These results lead to the conclusion that greedy
	flows existing with non-negligible probability contribute to the
	non-Gaussian nature of network traffic, because the decrease in
	$\alpha$ corresponds to an increase in the probability of
	a greedy flow existing and the increase in skewness corresponds to an
	increase in the degree of non-Gaussian nature of network traffic.

	\begin{figure}[tbp]
	   \setlength{\unitlength}{1mm}
	   \begin{center}
	   \begin{picture}(90,20)
	    \put(5,5){\circle{6}}
	    \put(4,4){\shortstack{C}}
	    \put(2,9){\shortstack{\scriptsize{TTL}\\\scriptsize{128}}}
	
	    \put(15,10){\circle{6}}
	    \put(14,9){\shortstack{R}}
	    \put(12,14){\shortstack{\scriptsize{TTL}\\\scriptsize{127}}}

	    \put(25,15){\circle{6}}
	    \put(24,14){\shortstack{R}}
	    \put(22,19){\shortstack{\scriptsize{TTL}\\\scriptsize{126}}}

	    \put(35,10){\circle{6}}
	    \put(34,9){\shortstack{R}}
	    \put(32,14){\shortstack{\scriptsize{TTL}\\\scriptsize{125}}}

	    \put(55,10){\circle{6}}
	    \put(54,9){\shortstack{R}}
	    \put(52,14){\shortstack{\scriptsize{TTL}\\\scriptsize{62}}}

	    \put(65,15){\circle{6}}
	    \put(64,14){\shortstack{R}}
	    \put(62,19){\shortstack{\scriptsize{TTL}\\\scriptsize{63}}}

	    \put(75,20){\circle{6}} 
	    \put(74,19){\shortstack{S}}
	    \put(72,11){\shortstack{\scriptsize{TTL}\\\scriptsize{64}}}

	    \put(8,6){\vector(2,1){5}}
	    \put(18,11){\vector(2,1){5}}
	    \put(28,14){\vector(2,-1){5}}

	    \put(38,10){\vector(1,0){5}}
	
	    \put(53,10){\vector(-1,0){5}}

	    \put(62,14){\vector(-2,-1){5}}
	    \put(72,19){\vector(-2,-1){5}}

	    \put(43,8){\framebox(5,3)}

	   \put(45.5,5){\line(0,1){3}}
	   \put(40,2){\shortstack{Measuring point}}
	   \end{picture}
	   \end{center}
	  \caption{Hop count estimation between two hosts.}
	  \label{fig:hop-count}
	  \end{figure}
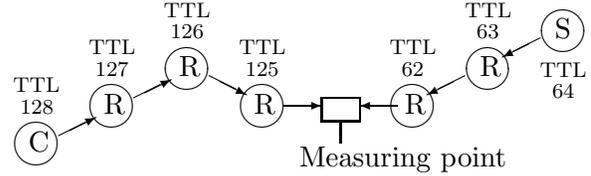

	\subsection{Hop count distribution of per-time-unit flow}
	For each per-time-unit flow $fl\_j\left(T_i\right)$, we
	investigated hop
	counts $hop\left(fl\_j\left(T_i\right)\right)$.
	We used only Data I because traces of Data II contain only
	one-way	traffic and we could not estimate hop counts of each
	flow with our method described below.
	To study hop counts between two nodes from the given trace
	data, we used the TTL (time to live) field of an IP packet.
	As its value is decreased when an IP packet passes a router,
	we can estimate hop counts between the source node and
	measuring point from the initial TTL value and the TTL value
	of the received packet.
	So, if we can obtain the hop counts from both the source and
	destination nodes to the measuring point, we can estimate the
	hop counts between these nodes.
	We show an example below.

	In Fig.~\ref{fig:hop-count}, client C and server S compose
	an IP flow. 
  	Let initial TTL values of each host to be 128 and 64.
	If we obtain TTL values as 125 and 62 at the measuring point,
	hop counts between these two nodes can be estimated as $\left(128
	- 125\right) + \left(64 - 62\right) + 1 = 6$, where we assume
	that the route between two nodes does not change during a
	round-trip.
	One difficulty with this approach is that the initial TTL
	values depend on the operating system or network equipment
	such as routers (see \cite{lance-99} for example).
	To overcome this difficulty, we employed the technique introduced
	by Fujii et al. in \cite{fujii-00}; that is, we assumed initial
	TTL values to be 32, 64, 128, and 255, and ignored other initial
	TTL	values such as 30 or 60 because systems that create such
	non-$2^n$ related initial TTL values can be considered to be
	rather out-of-date and unusual in today's network.
	From measured TTL values, we choose the closer (and larger) value
	from the above four values, and used it as its
	initial	TTL value.
	For example, if we receive a packet with initial TTL value
	of 45, we assume the initial TTL value of the packet to be 64.

	\begin{figure}[tbp]
	 \begin{center}
	\includegraphics[width=80mm]{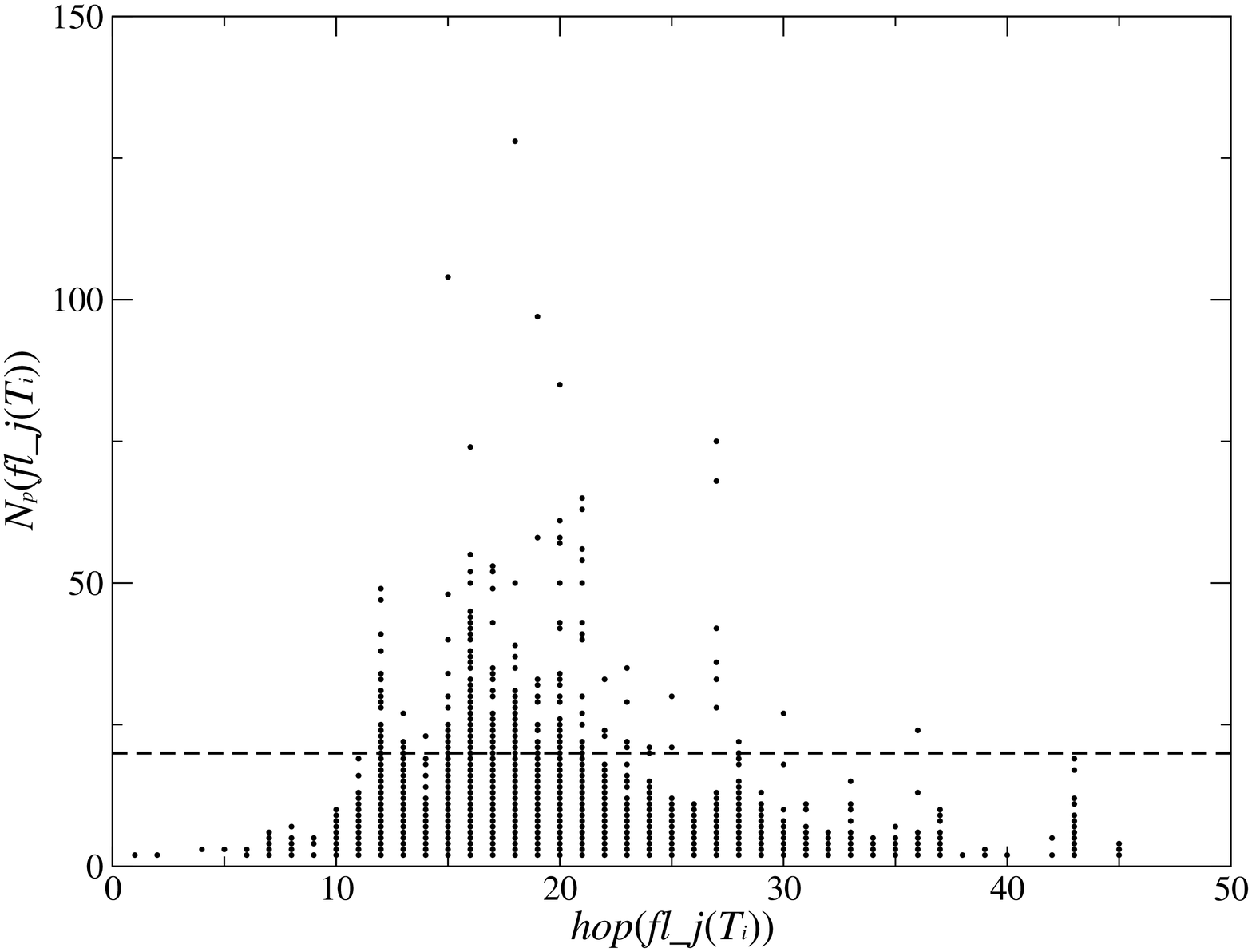}
	  \caption{$hop\left(fl\_j\left(T_i\right)\right)$
	 vs. $N_p\left(fl\_j\left(T_i\right)\right)$ for the trace
	of Data I given in Fig.~\ref{fig:non-Gauss-ex}.}
 	  \label{fig:ecl-hop-nump}
	 \end{center}
	\end{figure}

\begin{figure}[tbp]
	 \begin{center}
	\includegraphics[width=85mm]{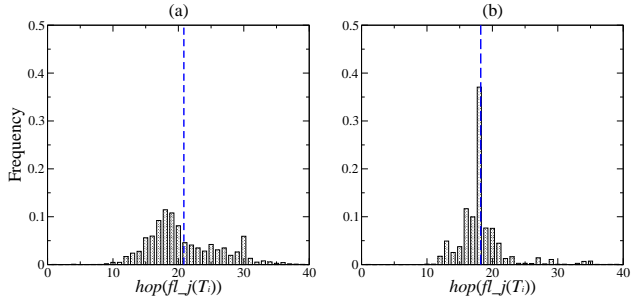}
	  \caption{Histogram of
	 $hop\left(fl\_j\left(T_i\right)\right)$ for all traces of
	 Data I. (a) is for all per-time-unit flows and (b) is for
	 greedy ones.
	The dashed lines indicate average hop counts.}
 	  \label{fig:hop-histo}
	 \end{center}
	\end{figure}

	We investigated hop counts $hop\left(fl\_j\left(T_i\right)\right)$
	of each per-time-unit flow	$fl\_j\left(T_i\right)$ for all traces
	of Data I,
	using the hop count estimation methodology described above.
	Here, we removed flows having multiple hop
	counts for one source IP address\footnote{This is caused by
	a change in routing or intentional change in initial TTL value
	given by some special applications such as traceroute.}.
	Figure~\ref{fig:ecl-hop-nump} shows the relationship between 
	$hop\left(fl\_j\left(T_i\right)\right)$ and 
	$N_p\left(fl\_j\left(T_i\right)\right)$ for the trace given in
	Fig~\ref{fig:non-Gauss-ex}.
	Hop counts of greedy flows (above the dashed line) can be
	considered to be smaller than those of all flows.
	Figure~\ref{fig:hop-histo} shows a histogram of
	$hop\left(fl\_j\left(T_i\right)\right)$ for (a) all
	per-time-unit flows and (b) for greedy per-time-unit flows,
	where we used all traces of Data I.
	Average hop counts for greedy flows were
	smaller than those of all flows, and most greedy flows had
	relatively smaller hop counts.	
	Actually, average hop counts were 20.84 for all flows and
	18.22 for greedy flows (see dashed lines in
	Fig.~\ref{fig:hop-histo})
	\footnote{
	In Fig.~\ref{fig:hop-histo}(b), the peak at hop count of
	18 is due mainly to one long-lived greedy IP flow, which
	caused a large number of greedy per-time-unit flows.
	Actually, the number of greedy per-time-unit flows coming from
	this IP flow was 7025, while total number of greedy	per-time-unit
	flows was 33817.  
	When we removed this IP flow, the average hop counts for greedy
	flows became 18.28, which indicates that the existence of the IP
	flow did not affect the result.
	}.

	From these results, we conclude that greedy flows had
	smaller hop counts than those of all flows in
	our study.
	This is because the RTTs of
	flows with smaller hop counts are assumed to be smaller
	statistically, as demonstrated in \cite{fujii-00}, and TCP flows
	with smaller RTTs can quickly make their window size larger
	(i.e., can become greedy) following the mechanism of TCP flow
	control.
	Accordingly we can assume that if TCP flows
	having smaller hop counts (i.e., smaller RTTs) exist with
	non-negligible probability, then the aggregated traffic shows
	a non-Gaussian nature.
	To verify the above assumptions, we examine the relationship between
	hop	counts and RTT, and protocol distribution for our trace data
	in the following two sections.		
	Our goal is to verify the linear relationship between hop counts
	and RTTs, and protocol breakdown for trace data used in this
	study.

	\subsection{Relationship between hop counts and RTTs}
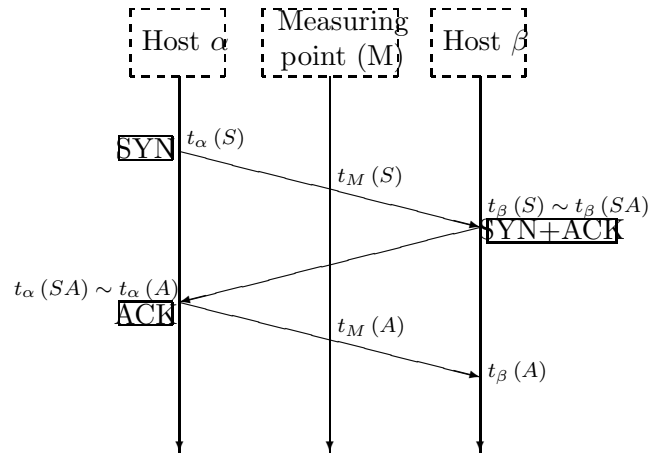
\begin{figure}[tbp]
	\setlength{\unitlength}{1mm}
		\begin{center}
		\begin{picture}(85,55)
		\put(23,50){\vector(0,-1){50}}
		\put(63,50){\vector(0,-1){50}}
		\put(43,50){\vector(0,-1){50}}

		\put(23,40){\vector(4,-1){40}}
		\put(63,30){\vector(-4,-1){40}}
		\put(23,20){\vector(4,-1){40}}

		\put(18,53.5){Host $\alpha$}
		\put(36,51.5){\shortstack{Measuring\\point (M)}}
		\put(58,53.5){Host $\beta$}

		\put(16.5,50){\dashbox(12.5,9)[c]}
		\put(34,50){\dashbox(18,9)[c]}
		\put(56.5,50){\dashbox(12.5,9)[c]}

		\put(15,39){\framebox(7,3)[c]{SYN}}
		\put(64,28){\framebox(17,3)[c]{SYN+ACK}}
		\put(15,17){\framebox(7,3)[c]{\shortstack{ACK}}}

		\put(24,41){\scriptsize{$t_{\alpha}\left(S\right)$}}
		\put(44,36){\scriptsize{$t_{M}\left(S\right)$}}
		\put(64,32){\scriptsize{$t_{\beta}\left(S\right)\sim
		t_{\beta}\left(SA\right)$}}

		\put(1,21){\scriptsize{$t_{\alpha}\left(SA\right)\sim
		t_{\alpha}\left(A\right)$}}
		\put(44,16){\scriptsize{$t_{M}\left(A\right)$}}
		\put(64,10){\scriptsize{$t_{\beta}\left(A\right)$}}


		\end{picture}
		\end{center}
	\caption{Diagram of TCP's 3-way handshake.}
	\label{fig:tcp-3way}
	\end{figure}

	Here, we introduce the technique for estimating RTTs from given
	passively measured traces and then give results of analyses for
	our trace data.
	Our approach is based on the technique proposed in
	\cite{fujii-00}, which is to analyze TCP's 3-way handshake
	\footnote{Basic 3-way handshake for connection synchronization
	is defined in RFC 793 \cite{rfc793}.} 
	packets.
	Adding to the description given in \cite{fujii-00}, we give a more
	detailed description of the technique.
	Figure~\ref{fig:tcp-3way} diagrams this between hosts $\alpha$
	and $\beta$.
	For convenience, we call a TCP packet with a SYN (SYN and ACK,
	ACK) flag bit on a SYN (SYN+ACK, ACK) packet.
	In our study, we measured traffic at measuring point M between
	the two hosts.

	First host $\alpha$ sends a SYN packet in order to request
	connection establishment.
	Let the time at this moment be $t_{\alpha}\left(S\right)$.
	The SYN packet passes the measuring point at
	$t_{M}\left(S\right)$, and is received by host
	$\beta$ at $t_\beta\left(S\right)$.
	Immediately upon receiving the SYN packet, host $\beta$ sends
	back a SYN+ACK packet at $t_\beta\left(SA\right)$, and it
	is received by host $\alpha$ at
	$t_\alpha\left(SA\right)$.
	Similarly, host $\alpha$ immediately sends back an ACK packet
	at $t_\alpha\left(A\right)$, and it passes measuring point M
	at $t_M\left(A\right)$, and it is received by host $\beta$ at
	$t_\beta\left(A\right)$ for the end of negotiation.
	Assuming that the delay caused by transactions of each host is
	quite small (i.e., $t_\beta\left(S\right)\sim
	t_\beta\left(SA\right),	t_\alpha\left(SA\right)\sim
	t_\alpha\left(A\right)$) and there is no queueing delay caused
	by network congestion, RTT between host $\alpha$ and $\beta$
	can be estimated as $t_M\left(A\right) - t_M\left(S\right)$ as
	described in the figure
	\footnote{Of course this assumption is not always appropriate;
	that is, conditions of host and network are always changing
	and RTTs for the same flows also fluctuate.
	So, to avoid the effect of fluctuations we used the average value
	of RTTs	for statistical study.}.

	Using the above approach, we estimated RTTs of IP flows that
	contain	TCP's 3-way handshake, where we removed flows that
	contain	duplicated SYN and SYN+ACK to estimate RTTs exactly.
	In the trace of Data I given in Fig.~\ref{fig:non-Gauss-ex},
	the total number of IP flows in the Internet-to-ECL
	direction was 16,358. We could estimate RTTs for 6834 of them
	and estimate both RTTs and hop counts for 1987 of them.

\begin{figure}[tbp]
	 \begin{center}
	\includegraphics[width=70mm]{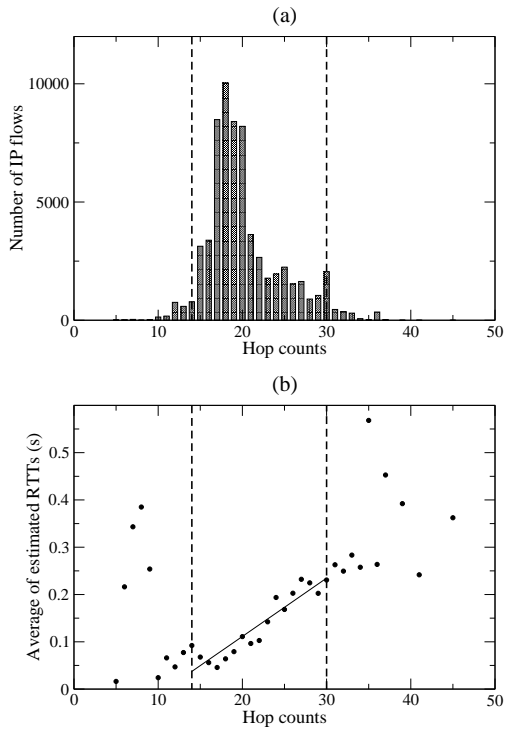}
	  \caption{(a) Hop counts vs. number of IP flows and (b) hop
	 counts vs. average of estimated RTTs.}
 	  \label{fig:ecl-hop-rtt}
	 \end{center}
	\end{figure}

	Figure~\ref{fig:ecl-hop-rtt} shows (a) the relationship between hop
	counts and number of IP flows and (b) the relationship
	between hop counts and the average of estimated RTTs, where we
	collected estimated RTTs per hop count and used their average
	as a representative value.
	The correlation coefficient of the average of the estimated
	RTTs and hop counts was 0.93, where the regression range of
	hop counts $h$ was set to $14 \leq h \leq 30$, where the
	number of IP flows exceeds 1\% (inside dashed lines). 
	The result indicates that hop counts and RTTs are in good
	agreement with a linear correlation.
	Thus, it was verified that statistically, IP flows with	smaller
	(larger) hop counts have smaller (larger) RTTs for trace data used
	in this study.
	This suggests that if the IP flow has smaller hop counts and its
	protocol is TCP, it can be greedier, following the mechanisms of
	TCP as mentioned in the previous paragraph.
	In the next section, we show the protocol distribution for trace
	data.

	\subsection{Protocol distribution}
	For each per-time-unit flow $fl\_j\left(T_i\right)$, we
	investigated the protocol.
	Table~\ref{tab:proto-all} shows the protocol distribution of
	all per-time-unit flows and greedy per-time-unit flows for
	Data I and II.
	The results show that for both cases, most of their protocols were
	TCP.
	That is, most per-time-unit flows followed TCP's flow control
	mechanism, and if flows had smaller RTTs (i.e., smaller hop
	counts as shown in previous section), they could quickly make
	their window size larger and be greedy in a short time,
	leading to the non-Gaussian nature of aggregated network traffic.
	As today's most popular Internet applications such as WWW are
	based on TCP (HTTP), this implication is very important.

\section{Conclusion}
	In this work, we showed the performance implication of the
	non-Gaussian nature of network traffic and analyzed its mechanisms
	using some Internet traffic data.
	Our main findings are that (1) the non-Gaussian nature degrades
	network performance, (2) it is caused by greedy existing
	flows with non-negligible probability, and (3) a large
	majority of greedy flows are TCP flows having relatively small hop
	counts, which correspond to small RTTs.
	Accordingly, we conclude that in a network that has greedy flows
	with non-negligible probability, a traffic controlling scheme or
	bandwidth design that considers non-Gaussian nature is essential.

	We expect that detecting non-Gaussian factors will
	allow us to propose practical methodologies for traffic engineering.
	We show some examples below.
	As we have shown in section 3, the behavior of each IP flow is
	related to its hop count; that is, IP flows with smaller hop
	counts tends to be greedier than ones with larger hop counts.
 	So, classifying IP flows with their hop counts (e.g., using TTL
	fields) at routers will be useful for traffic engineering.
	For instance, decreasing queueing priority for IP flows with
	smaller	hop counts will decrease the number of possible greedy
	flows.
	The decrease in the number of greedy flows will let the
	nature of aggregated traffic to be close to Gaussian, where
	network	performance will be improved for a given utilization and
	buffer capacity, as we showed in Fig.~\ref{fig:h-s-ploss}.
	It will also lead to the establishment of fairness among IP flows.
	Another example is to clarify the relationship between traffic
	characterization and network topology as mentioned in section 3.
	Efficient network design according to its topology will be
	established by this study.
	That is, from the hop count distribution obtained by
	analyzing the network topology, we can estimate whether the
	network	is likely to have greedy flows (i.e., IP flows
	having smaller hop counts).
	If it does with	non-negligible probability, then the aggregated
	traffic will show non-Gaussian nature and bandwidth design
	considering the effect of non-Gaussian nature will be	effective
	for efficient operation of the network.

        \begin{table}[tbp]
        \begin{center}
        \caption[]{Protocol distribution for all per-time-unit flows 
        and greedy ones.}
        \label{tab:proto-all}
                \begin{tabular}{p{10mm}|p{10mm}|p{10mm}|p{10mm}|p{10mm}}
                \hline
                \multicolumn{1}{c|}{} &
                \multicolumn{2}{c|}{Data I} &
                \multicolumn{2}{c}{Data II} \\
                \cline{2-5}
                \multicolumn{1}{c|}{} &
                \multicolumn{1}{c|}{All} &
                \multicolumn{1}{c|}{Greedy} &
                \multicolumn{1}{c|}{All} &
                \multicolumn{1}{c}{Greedy} \\
                \hline
                \multicolumn{1}{c|}{TCP} &
                \multicolumn{1}{c|}{97.00 \%} &
                \multicolumn{1}{c|}{99.90 \%} &
                \multicolumn{1}{c|}{95.76 \%} &
                \multicolumn{1}{c}{95.36 \%} \\
                \hline
                \multicolumn{1}{c|}{UDP} &
                \multicolumn{1}{c|}{1.02 \%} &
                \multicolumn{1}{c|}{0.10 \%} &
                \multicolumn{1}{c|}{4.20 \%} &
                \multicolumn{1}{c}{4.62 \%} \\
                \hline
                \multicolumn{1}{c|}{Other} &
                \multicolumn{1}{c|}{1.98 \%} &
                \multicolumn{1}{c|}{0.00 \%} &
                \multicolumn{1}{c|}{0.04 \%} &
                \multicolumn{1}{c}{0.01 \%} \\
                \hline
                \end{tabular}
        \end{center}
        \end{table}
 
 We consider that the existence of greedy flows is due to the heterogeneity
 of the Internet.
 As shown in section 3.2, flows were aggregated in various manners on
 a certain time scale.
 Actually, they were following a power-law
 (Fig.~\ref{fig:per-time-unit-numpkt-dist}).
 We expect that this power-law mainly comes from (a) the
 heterogeneity of network topology, which is partly found in
 hop count distributions or (b) the heterogeneity of user links,
 which ranges from low-speed links such as analog modem to high-speed
 links such as gigabit Ethernet.
 So, in the Internet, there exist various kinds of IP flows and this
 diversity leads to greedy flows existing with non-negligible
 probability.
 As pointed out in \cite{vern-97}, learning the characteristics of the
 Internet is an immensely challenging undertaking because of the
 network's great heterogeneity and rapid changes.
 However, we believe that seeking some invariant characteristics in
 the Internet such as self-similarity or non-Gaussian nature or
 power-law of per-time-unit flows will help us to build practical
 models of it and propose methodologies for operating it efficiently.

\end{document}